# Godel's Incompleteness Theorems and Platonic Metaphysics

Aleksandar Mikovic

Lusofona University, Lisbon

amikovic@ulusofona.pt

**Abstract**

We argue by using Godel's incompletness theorems in logic that platonism is the best metaphysics for science. This is based on the fact that a natural law in a platonic metaphysics represents a timeless order in the motion of matter, while a natural law in a materialistic metaphysics can be only defined as a temporary order which appears at random in the chaotic motion of matter. Although a logical possibility, one can argue that this type of metaphysics is highly implausible. Given that mathematics fits naturally within platonism, we conclude that a platonic metaphysics is more preferable than a materialistic metaphysics.

**1. Introduction**

Many physicists beleive in a materialistic metaphysics, which means that the basic elements of the reality are space, time, elementary particles (in the case of string theory, one can replace the elementary particles with strings) and the corresponding fields. There is also a finite set of dynamical laws which describe the motion of the elementary particles and the fields. These laws can be expressed mathematically, so that one arrives at a mathematical theory of everything (TOE). String theory is the famous example of such a metaphysics. On the other hand, many mathematicans are platonists, which means that they beleive that the mathematical ideas are eternal truths, which exist outside of space and time. Since all known physics laws can be described by mathematics, then it is plausible to propose a platonic metaphysics where our universe is the same object as the mathematical structure of the ultimate TOE [1]. This is the modern-day version of the pythagorean doctrine that all things are numbers.

Tegmark's mathematical universe metaphysics proposed in [1] was criticised in [2], where it was argued that the concept of passage of time is inherently non-mathematical. Hence a platonic metaphysics has to include the idea of time flow in order to better describe the reality. The basic idea of [2] is that our universe is a block universe, containing both mathematical and non-mathematical structures, with a moving three-dimensional Cauchy surface (a burning block universe). This Cauchy surface splits the block universe into the past, the present and the future.

A block universe without the time flow is simply an abstract object in the platonic realm of ideas. Any TOE one can write down represents an object of this type. A real universe is a mathematical structure in time. Hence a temporal platonic metaphysics can serve as a good metaphysics for explaining both mathematics and physics.

However, there is an obstruction to the TOE concept, which comes from the Goedel's incompletness theorems (GIT) in logic, see [3]. The first incompletness theorem states that a finite or recursive logical system, which includes arithmetics, has a finite demonstrational power. More precisely, one can construct an undecidable statement (the Godel statement) which can not be proven to be true or false by using the postulates of the theory. Furthermore, the Godel statement can be taken to be true and therefore can be added to the initial set of the postulates. In the new theory there will also exist a Godel statement, and so on, ad infinitum. If we apply this theorem to a TOE, it means that there will exist statements, i.e. physical phenomena, that cannot be explained within this theory. The first person to point out this was S. Jaki [2] in the 60's, and later it was also discussed by S. Hawking [3]. Although GIT prevent the existence of a TOE, the concept of a TOE is still useful, since one would like to have a mathematical theory which describes a sufficiently large number of physical phenomena. In modern physics this is reflected by the ongoing effort to find a mathematical theory which unifies Quantum Mechanics and General Relativity and explains the Standard Model parameters.

The second theorem is more technical, and states that a consistent theory T cannot have a statement referring to the consistency of T. It serves to stregthen the first theorem, since a Godel statement does not refer to the consistency of the theory T, see [3].

There is another consequence of GITs that we would like to elaborate here. We will argue that GITs favor a platonic metaphysics.

## 2. Metaphysics and the laws of Nature

A law of Nature can be easily understood within a platonic metaphysics. It is a postulate in a mathematical theory we use to describe the Nature. On the other hand, explaining the laws of Nature within a materialistic metaphysics, is more complicated. If one accepts that the natural laws are different entities from space, time and matter, and are irreducible, then for a materialist it does not seem to be a problem to add a finite set of such objects to his ontology. In this case the natural laws are the postulates of a finite TOE. However, any TOE has to include the aritmethics, so that Goedel's theorems imply that there can not be a finite number of laws which completely explain the universe and one must introduce an infinite number of natural laws. This means that in addition to space, time and matter, one has to introduce an infinite number of other entities, which are not reducible to space, time and matter, and hence one is back at platonism.

In order to avoid introducing an infinite number of non-material entities in a materialistic metaphysics, one then has to give up the idea of a natural law as a mathematical concept (i.e. a postulate in a mathematical theory). Then the only explanation for a natural law in a materialistic metaphysics is that a natural law represents a regular pattern which appears in the fundamentaly chaotic motion of matter. This regularity appears at random and lasts for a very long time. In this case one accepts the view that at the fundamental level there is no order and the particle trajectories and field configurations are completely arbitrary. This doctrine is a logical possibility, but it is highly implausible. The standard example for this type of implausibility is to find a string of letters in a random sequence of letters which corresponds to a well-known novel; or to construct a functioning airplane by using a tornado passing through a junk yard. Also, if the natural laws are finite-duration random regularities, then the Earth can stop orbiting the Sun tomorrow, which means that our reality can disintegrate at anytime in the future.

Another problem in a materialistic metaphysics is how an observer will recognize a natural law given that the ideas of order do not exist. This is also a problem in a platonic metaphysics, see [6], but it is a less severe problem, because the basic elements from which one can construct a solution already exist, see [2] for a possible solution.

## 3. Conclusions

Since in the spirit of science is the belief that the natural laws can be described by mathematics, than platonism is a natural metaphysics for science. The natural laws must be mathematical, because by definition, a natural law must be expressible by a finite string of symbols obeying the laws of logic. One can also have a natural law within a materialistic metaphysics, but its meaning is completely different from the meaninig of a natural law in a platonic metaphysics. While a natural law in a platonic metaphysics represents a timeless order in the motion of matter, in a materialistic metaphysics a natural law is a temporary order which appears in the chaotic motion of matter at random and lasts for a long time. Although a logical possibility, one can argue that this scenario is highly implausible. Hence a platonic metaphysics is more plausible than a materialistic metaphysics.

A platonic metaphysics can also explain the famous Wigner observation about the unreasonable effectivnes of mathematics in science. Our universe can be thought of as a block universe containing the mathematical structures which represent the natural laws. However, the block-universe picture is not complete without the idea of passage of time, which can be implemented by using a temporal platonic metaphysics [2]. Therefore a platonic metaphysics has a greater explanatory power than a materialistic metaphysics.

Another important feature of a platonic metaphysics is that the mathematical ideas alone are not sufficient to describe the real world. For example, the idea of passage of time is not mathematical, i.e. can not be completely described by a finite set of symbols, or in other words, by a finite text. It can be grasped only intuitively. This also applies to other important non-mathematical ideas, such as conscioussnes and the idea of an observer. Godel's theorems are consistent with the fact that we perceive certain abstract ideas as real, although there is no formal definition or a proof of those.